# Prediction of a spin-polarized two-dimensional electron gas at the LaAlO$_3$/EuO(001) interface

Yong Wang, Manish K. Niranjan, J. D. Burton, Joonhee M. An,
Kirill D. Belashchenko, and Evgeny Y. Tsymbal[*]

*Department of Physics and Astronomy, Nebraska Center for Materials and Nanoscience,
University of Nebraska, Lincoln, Nebraska 68588, USA*

First-principles calculations predict the existence of a spin-polarized two-dimensional electron gas (2DEG) at the LaO/EuO interface in a LaAlO$_3$/EuO (001) heterostructure. This polar interface favors electron doping into the Eu-$5d$ conduction bands resulting in a 2DEG formed at the interface. Due to the exchange splitting of the Eu-$5d$ states the 2DEG is spin-polarized below the Curie temperature of EuO. The predicted mechanism for the formation of a spin-polarized 2DEG at the interface between polar and ferromagnetic insulators may provide a robust magnetism of the 2DEG which is interesting for spintronics applications.

The field of research related to complex oxide materials has been developing vigorously in the last few decades. Oxides exhibit an abundance of macroscopic physical properties, often involving the interplay between magnetism, ferroelectricity, and conductivity.[1] Even a more rich spectrum of physical phenomena occurs if two or more oxides are combined with atomic-scale precision in a heterostructure to form a new nanoscale material.[2,3] Recent advances in thin-film deposition and characterization techniques made possible the experimental realization of such oxide heterostructures, promising novel functionalities and device concepts. A prominent example is the recent discovery of the formation of a metallic phase at the interface between two oxide insulators, SrTiO$_3$ and LaAlO$_3$.[4] It was found that this metallic phase is confined within a couple of nanometers near the interface[5] and therefore can be regarded as a two dimensional electron gas (2DEG). The 2DEG has a very high carrier density and a relatively high carrier mobility making it attractive for applications in nanoelectronics, e.g., as all-oxide field-effect transistors.[6,7] These properties of the 2DEG at oxide interfaces have stimulated significant research activity both in experiment[8-13] and in theory.[14-20]

Recently it was found that at ultra-low temperatures the 2DEG occurring at the interface between the non-magnetic LaAlO$_3$ and SrTiO$_3$ materials may become magnetic.[21] This behavior was attributed to the exchange splitting of the induced electrons in the Ti-$3d$ conduction band, which is corroborated by spin-polarized first principles calculations of LaAlO$_3$/SrTiO$_3$,[18,19] as well as LaTiO$_3$/SrTiO$_3$ interfaces.[16] Making 2DEG gas spin-polarized is a very exciting prospect for spintronics applications, where the involvement of the spin degree of freedom broadens the spectrum of potential applications.[22] Very recently it was proposed that it may be possible to create a fully spin-polarized 2D electron gas by replacing one monolayer of SrO by LaO in SrMnO$_3$.[23]

Here we pursue a different route to achieve a spin-polarized two-dimensional electron gas, by employing a ferromagnetic insulator as one of the constituents in the oxide heterostructure. Spin-polarized properties of the 2DEG are, in this case, expected to be inherited from the ferromagnetism of the oxide, and consequently this approach may lead to a more robust magnetism in the 2DEG, which is beneficial for applications. To illustrate the idea we consider EuO as a representative ferromagnetic insulator in conjunction with LaAlO$_3$ to form a spin-polarized 2DEG at the LaAlO$_3$/EuO(001) interface.

EuO has a rocksalt crystal structure and is a ferromagnetic insulator (semiconductor) with a bulk Curie temperature ($T_C$) of 69K. A divalent Eu ion in EuO has a half-filled $4f$ shell leading to the $^8S_{7/2}$ ground state and the magnetic moment of $7\mu_B$ per Eu ion. The Heisenberg exchange coupling between the localized $4f$ electrons causes the ferromagnetic ordering in EuO below $T_C$. The half-filled $4f$ band is separated from the $5d$-$6s$ conduction bands by a bandgap of 1.12 eV at room temperature.[24] In the ferromagnetic state of EuO, the direct exchange interaction between the localized $4f$ moments and the delocalized $5d$ conduction band states leads to the spin splitting of the latter. The spin splitting of the $5d$ states as large as 0.6eV produces the full spin polarization near the bottom of conduction band.[25] This feature of the EuO conduction band allows creating a highly spin-polarized electron gas by appropriate n-doping of the material. In particular, recently it was found that the transport spin polarization of conduction electrons in EuO doped with La exceeds 90%.[26]

Here we demonstrate that the electron doping may be achieved locally at the LaAlO$_3$/EuO(001) interface to create a spin-polarized 2DEG. The mechanism responsible for the 2DEG formation is similar to that known for the LaAlO$_3$/SrTiO$_3$ interface.[27] LaAlO$_3$ consists of alternating $(LaO)^+$ and $(AlO_2)^-$ charged planes, whereas EuO consists of $(EuO)^0$ neutral planes. When LaAlO$_3$ is deposited on top of EuO the divergence in the electrostatic potential can be avoided by transferring half an electron per two-dimensional unit cell to the LaO/EuO terminated interface. A charge transfer to the interface also occurs if the LaAlO$_3$ layer is non-stoichiometric and terminated with the LaO monolayers on both sides. In this case an "extra" electron is introduced into the system due to the uncompensated ionic charge on the additional $(LaO)^+$ monolayer. This electronic charge is accommodated by partially occupying conduction band states near the interface, producing a 2DEG. In ferromagnetic EuO the conduction band is formed by exchange split Eu-5$d$ states and so the 2DEG is expected to be spin-polarized. Thus, by forming the LaAlO$_3$/EuO(001) interface one can achieve a spin-polarized 2DEG below the Curie temperature of EuO.

To quantitatively demonstrate our prediction we perform first-principles calculations of the electronic structure of the LaAlO$_3$/EuO (001) interface within the framework of density functional theory (DFT)[28] implemented within Vienna *Ab-Initio* Simulation Package (VASP).[29] The spin-polarized calculations include the electron-ion potential which is described within the projected augmented method (PAW).[30] The exchange-correlation effects are treated within the Perdew-Burke-Ernzerhoff (PBE) form[31] of the generalized gradient approximation (GGA). Self-consistent calculations are performed using a plane-wave basis set limited by a cutoff energy of 520eV and the 6×6×1 Monkhorst-Pack k-point mesh[32] with energy converged to $10^{-5}$ eV/cell. Atomic relaxations are performed until the Hellmann-Feynman forces on atoms have become less than 30 meV/Å.

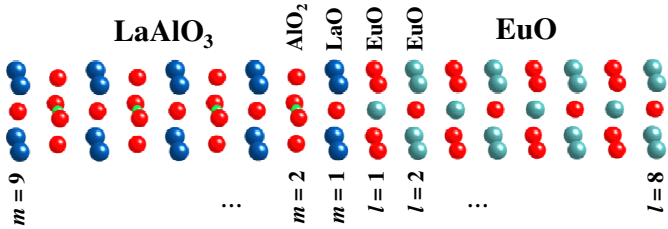

**Fig. 1**: Atomic structure of the LaO/EuO interface in the (LaAlO$_3$)$_{8.5}$/(EuO)$_{15}$ (001) superlattice containing 8.5 unit cells of LaAlO$_3$ and 15 monolayers of EuO within the supercell. Indices $l$ and $m$ denote atomic monolayers and are increasing with separation from the interface. $l = 8$ in EuO and $m = 9$ in LaAlO$_3$ correspond the middle of the respective layers.

On-site correlations for the Eu-4$f$ orbitals are included within the GGA+$U$ approach.[33] The value of $J = 0.6$ eV is calculated using the constrained occupation method[34] by considering the 4$f$ states as an open-core shell and finding the GGA energy difference between the $4f_\uparrow^7 4f_\downarrow^0$ and $4f_\uparrow^6 4f_\downarrow^1$ configurations.[35] We find, however, that the value of $U = 5.3$ eV obtained by this method appears to be too small and leads to the 4$f$ states being too shallow with respect to the conduction band of EuO. This discrepancy is due to the underestimation by GGA of the intrinsic insulating gap between the O-2$p$ and Eu-5$d$ states. Therefore, we adjusted the value of $U$ empirically and found that $U = 7.5$ eV results in a very reasonable agreement with experiment. In particular, $U + 6J = 11.1$ eV agrees well with the occupied-unoccupied 4$f$ state splitting observed in photoemission-inverse photoemission for Eu metal; the optical band gap at the X point of 0.94 eV is consistent with the zero-temperature experimental value of approximately 0.95 eV;[36] and the lattice constant $a = 5.188$ Å agrees with the experimental value of $a = 5.144$ Å. The exchange splitting of the Eu 5$d$ orbitals is found to be $\Delta_d = 0.75$ eV. Since the La-4$f$ bands lie at higher energy than that predicted by GGA, we impose $U = 11$ eV on these orbitals to avoid their spurious mixing with the conduction bands of LaAlO$_3$.[16]

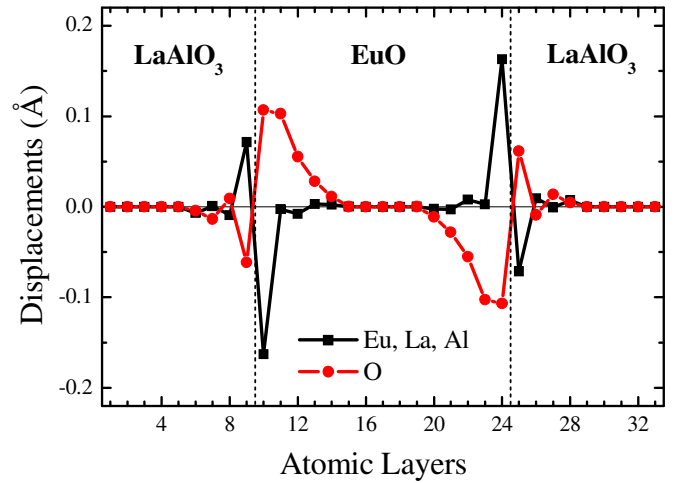

**Fig. 2**: Atomic displacements in the (LaAlO$_3$)$_{8.5}$/(EuO)$_{15}$ (001) superlattice with respect to the atomic "bulk" positions. The latter are determined by fixing the in-plane lattice constant and optimizing the interlayer distance, as described in text. The vertical dashed lines indicate interfaces.

We consider the LaO/EuO terminated interface of a (LaAlO$_3$)$_{8.5}$/(EuO)$_{15}$ superlattice (i.e., a supercell containing 8.5 unit cells of LaAlO$_3$ and 15 monolayers of EuO) stacked in the [001] direction, as shown in Fig.1. We use periodic boundary conditions and impose mirror plane symmetry at the central EuO monolayer. The in-plane lattice constant of the superlattice is fixed to the calculated lattice constant of cubic LaAlO$_3$, $a = 3.81$ Å, which is in good agreement with the experimental value $a = 3.79$ Å. Under this constraint EuO exhibits tetragonal distortion of $c/a = 0.947$. This distortion does not change significantly the electronic structure of bulk EuO. In particular, the band gap of EuO becomes $E_g = 1.15$ eV and the exchange splitting of the $d$ orbitals becomes $\Delta_d = 0.66$eV. The out-of-plane lattice constant of the supercell is determined by optimizing the interface separation distance between the LaAlO$_3$ and EuO sub-units keeping their bulk lattice constants fixed. Then, we fix the supercell dimensions and relax the atomic positions of all the atoms in the (LaAlO$_3$)$_{8.5}$/(EuO)$_{15}$ superlattice.

Fig. 2 shows the calculated atomic displacements within the (LaAlO$_3$)$_{8.5}$/(EuO)$_{15}$ supercell. It is seen that the largest structural relaxations occur in the vicinity of the LaO/EuO interfaces and involve a polar distortion in which the negatively charged O anions are displaced with respect to the positively charged cations (either La, Al or Eu). When moving away from the interface, the magnitude of the displacements decays and the La-La and Eu-Eu distances revert to constant values close to those in the bulk.

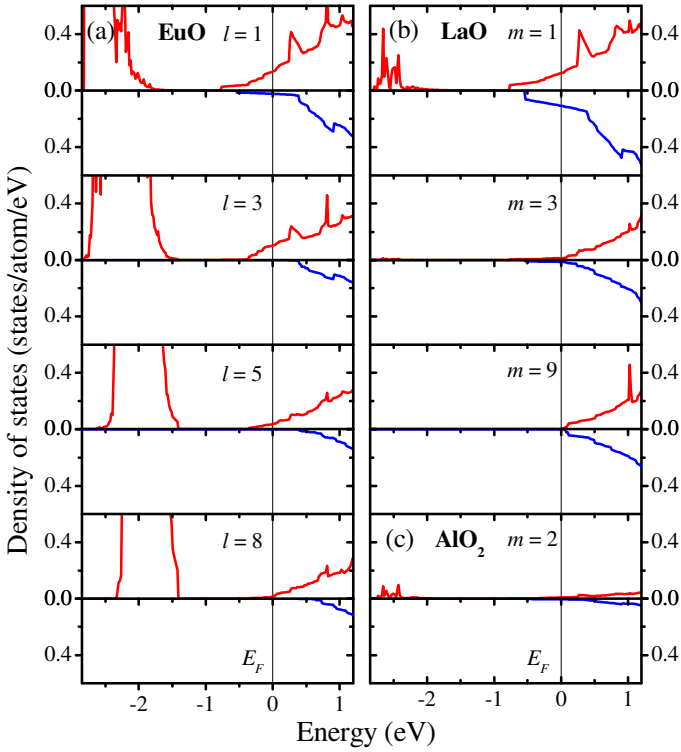

**Fig. 3**: Layer- and spin-resolved density of states (DOS) on EuO (a) LaO (b) and AlO$_2$ (c) monolayers located at different planes *l* and *m* away from the LaO/EuO interface (as labeled in Fig.1). Top (bottom) panels show the majority(minority)-spin. In panel (a) the majority-spin states at energies below –1eV are the occupied Eu-4*f* states. The vertical lines denote the Fermi energy ($E_F$).

Fig. 3 shows the layer resolved majority- and minority-spin densities of states (DOS) on the EuO (a) and LaO (b) monolayers located at different planes *l* and *m* away from the LaO/EuO interface (see labeling in Fig.1). It is seen that there are occupied states below the Fermi energy on both the LaO and EuO monolayers near the interface which indicate the formation of the n-type 2DEG at the LaO/EuO interface. The DOS on AlO$_2$ monolayers (Fig. 3c) is negligible near the

Fermi energy compared to that on the LaO and EuO monolayers. Far away from the interface, the DOS at the Fermi energy on both the LaO and EuO monolayers drops to zero reflecting the insulating nature of bulk LaAlO$_3$ and EuO.

The central result of our calculation is the formation of spin-polarized 2DEG at the LaAlO$_3$/EuO(001) interface. This fact is evident from Fig. 3a indicating a significant difference in the occupation of the EuO majority- and minority-spin conduction bands near the interface. This is the direct consequence of the exchange splitting of the conduction 5*d* states in EuO and the occupation of these states due to the electron doping of the interface. As seen from Fig. 3a, starting from the third EuO monolayer (*l* = 3) away from the LaO/EuO interface only majority-spin states are occupied in the conduction band of EuO indicating tendency to half-metallicity known for the n-doped bulk EuO.[25] Figs. 3b reveals the spin splitting of the conduction bands of LaAlO$_3$ which is the result of the exchange interaction between the Eu-4*f* and La-5*d* states across the interface.

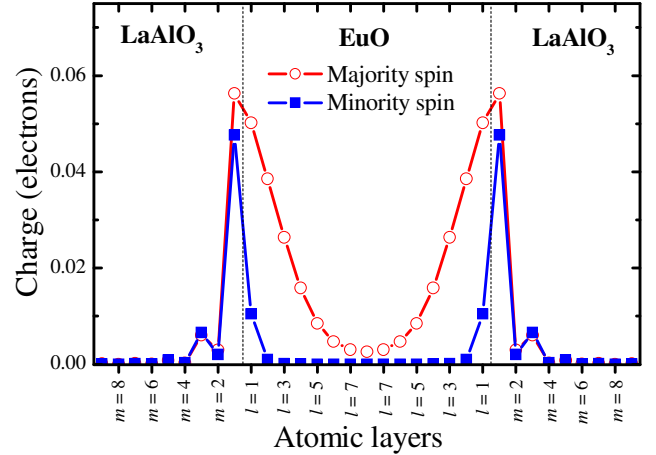

**Fig. 4**: Spin-dependent charge distribution across the (LaAlO$_3$)$_{8.5}$/(EuO)$_{15}$ (001) supercell. The notation for the atomic layers is the same as in Fig.1. The dashed lines indicate interfaces.

For the LaAlO$_3$/EuO interface we find that the top of the O-2*p* valence bands lies at about –3.6eV in LaAlO$_3$ and at about –3.0eV in EuO with respect to the Fermi energy. Due to a much larger energy gap between the O-*p* valence bands and the conduction bands in LaAlO$_3$ than in EuO (the calculated values are 3.7eV and 2.7eV respectively and the experimental values are 5.6eV and 4.4eV[37] respectively), the conduction band minimum lies lower in EuO than in LaAlO$_3$ resulting in the charge accumulating mainly within the EuO layer. This is evident from Fig. 4 which shows the distribution of the spin-dependent charge across the unit cell. These charges are calculated by integrating the spin- and layer-resolved DOS from the conduction band minimum up to the Fermi energy.[38] The large spin polarization of the 2DEG is seen at the EuO monolayers near the LaAlO$_3$/EuO interface. The estimated value of the spin polarization of the free charge density is about 50%.

We note that our calculation is performed for the non-stoichiometric $LaAlO_3$ layer which is assumed to be LaO terminated on both sides. Similar to the previous theoretical studies,[14-20] in this geometry an "extra" electron is introduced in the system due to the uncompensated ionic charge on the additional $(LaO)^+$ monolayer. For a stoichiometric $LaAlO_3$ layer deposited on top of EuO we expect the formation of 2DEG due to electronic reconstruction resulting from the charge transfer to the interface that eliminates the increasing electrostatic potential in $LaAlO_3$.[29] Due to dissimilar band alignments at the $LaAlO_3$/EuO interface compared to the $LaAlO_3$/$SrTiO_3$ interface we anticipate a different critical thickness of $LaAlO_3$ required to form a 2DEG.[6]

In summary, based on first-principles calculations we have predicted the possibility to create a spin-polarized 2DEG at the LaO/EuO interface in the $LaAlO_3$/EuO (001) heterostructure. We demonstrated that this polar interface favors electron doping into the Eu-$5d$ conduction bands rendering a 2DEG formed at the interface. Due to the exchange splitting of the Eu-$5d$ states the 2DEG becomes spin-polarized below the Curie temperature of EuO. The predicted mechanism for the formation of a spin-polarized 2DEG at the interface between polar and ferromagnetic insulators may lead to a robust magnetism of a 2DEG which is interesting for spintronics applications.

This work was supported by the National Science Foundation through the Materials Science Research and Engineering Center (NSF-DMR Grant No. 0820521), the Nanoelectronics Research Initiative of Semiconductor Research Corporation, and the Nebraska Research Initiative. Computations were performed at the Research Computing Facility of the University of Nebraska-Lincoln and the Center for Nanophase Materials Sciences at Oak Ridge National Laboratory.

*e-mail: tsymbal@unl.edu